\documentclass[11pt]{article}
\usepackage{amsfonts}

\let\ssection=\section
\renewcommand{\section}{\setcounter{equation}{0}\ssection}
\newfam\eusmfam
\newfam\scrfam
\batchmode\font\tenscr=rsfs10 scaled \magstep1 \errorstopmode
\ifx\tenscr\\else	\font\sevenscr=rsfs7 scaled \magstep1
	\font\fivescr=rsfs5 scaled \magstep1
	\skewchar\tenscr='177 \skewchar\sevenscr='177 \skewchar\fivescr='177
	\textfont\scrfam=\tenscr \scriptfont\scrfam=\sevenscr
	\scriptscriptfont\scrfam=\fivescr
	\def\scr{\fam\scrfam}
	\def\cal{\scr}
\fi

\newcommand{\eqn}[1]{\label{#1}}
\newcommand{\eq}[1]{\begin{equation}#1\end{equation}}

\newcommand{\eqs}[1]{\begin{eqnarray}#1\end{eqnarray}}

\newcommand{\bra}[3]{{}_{#2}^{#3}\big{<}#1\big{|}}
\newcommand{\ket}[3]{\big{|}#1\big{>}_{#2}^{#3} }

\newcommand{\vev}[2]{{\big <} #1 {\big >}_{#2}}
\def\lb{\nonumber\\}
\newcommand{\refbr}[1]{(\ref{#1})}

\def\a{\alpha}

\def\g{\gamma}
\def\d{\delta}
\def\e{\epsilon}

\def\F{\Phi}

\def\k{\kappa}

\def\m{\mu}
\def\n{\nu}
\def\r{\rho}
\def\p{\pi}
\def\P{\Pi}
\def\s{\sigma}
\def\S{\Sigma}
\def\th{\theta}

\def\G{\Gamma}
\def\D{\Delta}

\def\O{\Omega}
\def\L{\Lambda}
\def\/{\over}
\def\*{\partial}
\def\|{\mid}
\def\barz{\bar{z}}

\def\2{\half}
\def\3{{1\/3}}

\def\la{\leftarrow}
\def\ra{\rightarrow}
\def\na{\nabla}

\renewcommand{\baselinestretch}{1.5}
\newcommand{\NP}[1]{Nucl. Phys.\ {\bf #1}\ }

\newcommand{\CMP}[1]{Comm. Math. Phys.\ {\bf #1}\ }

\newcommand{\PR}[1]{Phys. Rev.\ {\bf #1}\ }

\newcommand{\IJMP}[1]{Int. J. Mod. Phys.\ {\bf #1} \ }

\newcommand{\AP}[1]{Ann. Phys.\ {\bf #1} \ }

\newcommand{\half}{\mbox{\scriptsize $ \frac{1}{2}$}}

\newcommand{\HH}{\ensuremath{\mathcal H}}

\newcommand{\CC}{\ensuremath{\mathcal C}}

\begin{document}

\newpage
\vspace*{-3cm}
\pagenumbering{arabic}
\begin{flushleft}
G\"{o}teborg\\
ITP 96-5\\
hep-th/9604010\\
April 1996
\end{flushleft}
\vspace{12mm}
\begin{center}
{\huge Finite Deformations of Conformal Field Theories Using
Analytically Regularized Connections}\\[14mm]
\renewcommand{\baselinestretch}{1.2}
\renewcommand{\footnotesep}{10pt}
{\large Alexander von Gussich\footnote{tfeavg@fy.chalmers.se, 
tel: +46 31 772 3159, fax: +46 31 772 3204}\\
Per Sundell\footnote{tfepsu@fy.chalmers.se, tel: +46 31 772 3156, 
fax: +46 31 772 3204}\\
}
\vspace{12mm}
{ Institute of Theoretical Physics\\Chalmers University of Technology\\ and
University of G\"{o}teborg\\
S-412 96 G\"{o}teborg, Sweden}
\end{center}
\vspace{3mm}
\begin{abstract}
We study some natural connections on spaces of conformal field theories  
using an analytical regularization method. 
The connections are based on marginal conformal field theory deformations. 
We show that the analytical regularization preserves conformal invariance and 
leads to integrability of the marginal deformations.
The connections are shown to be flat and to generate well-defined
finite parallel transport.
These finite parallel transports yield formulations of the deformed theories 
in the state space of an undeformed theory. The restrictions of 
the connections to the tangent space are curved but free of torsion. 
\end{abstract}
\renewcommand{\baselinestretch}{1.5}
\vfill
\newpage

\section{Introduction and Summary}

In string theory conformal field theories are the classical solutions
\cite{Callan}.
We know how to formulate the perturbation theory around such
backgrounds, either as a ``sum-over-surface'' expansion of on-shell amplitudes or as
off-shell string field theory. 
The question is to what extent these perturbative
formulations can be made background independent \cite{Sen}.
In string field theory one has the opportunity of studying 
the perturbations of a conformal background
in a generic off-shell direction. Such deformations point to 
2D field theories which are not conformal.
Nevertheless one indeed has formal invariance of the underlying algebraic structures of
the string field theory. We however think that there might still
be some issues to raise concerning the regularization of such deformations
such that the deformed formulation is really a regular theory, 
say of some perturbed non-conformal theory.

This issue has been studied in more detail in the case that the perturbation 
points in a tangent direction to the space of classical solutions. One can then use the 
on-shell formulation of string theory given in terms of conformal field theory.
These on-shell deformations can be described by insertions of marginal perturbations
$\int_{\S}d^{2}z\hat{\F}(z,\bar{z})$ into the correlation functions on $\S$.
Here the operator $\hat{\F}(z,\bar{z})$ has conformal weight $(1,1)$ w.r.t. 
stress-energy tensor of the unperturbed conformal field theory on $\S$. 
Such marginal perturbations preserves the conformal invariance, with the new stress-energy 
tensor given by $\hat{T}'(z,\bar{z})=\hat{T}(z)+\frac{\bar{z}}{z}\hat{\F}(z,\bar{z})$. 
The problem is that the marginal operator may not remain marginal under finite 
transformations, which leads to non-integrability of these deformations. 
That's no disaster
but integrability is desired since it will lead to that one can obtain a finite 
parameterization of some subspace of the space of classical solutions. Of course
this must be extremely  
complicated and by marginal perturbations we only expect to probe a small
``branch'' of this space.

Marginal deformations of conformal field theories (CFT's) have been studied at numerous 
occasions \cite{Kadanoff,Freericks,Evans,Campell}. In \cite{cvgs} marginal 
deformations 
were studied using a particular regularization method based on 
analytical continuation. There has been question both about the freedom to
choose regularization and what it should look like \cite{Ranganathan,RangSon,Pelts}. 
Regularization by analytical continuation has also been considered in four dimensions, 
and has been shown to be equivalent to dimensional regularization \cite{Hawking}.
The need of a regularization is due to the fact that
singularities occur when integrals of the marginal operators are inserted
into the vacuum correlation functions in order to perturb them.
In general this regularization may ruin the conformal invariance of the
deformed theory as well as integrability.
In \cite{cvgs} it was shown that the analytical 
regularization method indeed preserves conformal invariance,  
and the method was adopted in the construction of a
particular connection $\d$. 

In section 2 and 3 we will explain the analytical regularization method and
in section 4 and 5 we will explain the role of the connections in the 
study of marginal deformations of CFT's within the operator formalism.
In section 6 the analytical regularization method is used
to construct a one parameter family of natural connections
$\nabla^{(R)}$, $0< R\leq 1$, and discuss their connection to the connection $\d$. 
The connections $\nabla^{(R)}$ for two radii $R>R'$ are related
by a tensor which is the integral over the local annulus $R>|z|>R'$ of the marginal 
operator
that is used to construct the connections. 
This tensor is shown to generate inner transformations
on the formulations of the perturbed theories in the unperturbed
state space, corresponding to changes of $R$.
The relation between the connections $\d$ and $\nabla^{(R)}$ is explained in section 7.
We would like to stress that the parameter $R$ does
not play the role of any regularization. That is, the discs
of radius $R$ are not cut out of the correlators and treated separately.
As we will see, the choice of
$R$ merely reflects the arbitrariness
in the choice of the size of the local coordinate discs associated with the 
external states in a correlation function.
By a natural connection we mean a connection that can be defined using only the data of the
conformal field theory. The bases dependence (in case there are no preferred bases)
of such natural connections will be discussed in section 6.

The connection $\nabla^{(1)}$ and the connection $\d$ are compatible with the Zamolodchikov
metric, used in the definition of the sewing operation.
However, as a consequence of
the analytical regularization method all the connections $\nabla^{(R)}$ (and $\d$) 
commute with the
sewing operation for all values of $R$. Hence the connections generate
(infinitesimal) conformal field theory deformations. This will be explained 
in section 9.
Before we can do this we have to discuss the deformations of the 
Virasoro algebra and the
integrability of the marginal operators 
and demonstrate how higher powers of $\nabla^{(R)}$ 
generate regular finite conformal field theory deformations. This is done in section 
7 and 8.
This means that a finite space of string backgrounds can be represented in the state space
of a fixed background. 
This result supports the results on background invariance of string 
field theory referred to above.

The connections $\nabla^{(R)}$ and $\d$ are 
flat on the full state space of the conformal field
theories, so the finite parallel transport is path independent, while their
restrictions to the physical subspace acquire induced curvature
but no torsion (the induced curvature vanishes for 
$\d$).  Whether or not the connections has any meaning as local restrictions
of globally defined objects on a ``CFT-manifold'' is not clear to us. 
Of course the connection $\d$ seem to be too simple, while for
$R>0$ the connections at least possess some non-trivial geometrical structure
on the physical subspace.

\section{The Operator Formalism}

In this section we will state some of the basic properties of the operator 
formalism \cite{Vafa}.
Properties that will be frequently used in the remainder of this paper.

In the operator formalism the CFT takes an arbitrary compact genus g 
Riemann surface
$\S$ and maps it onto a surface state $\bra{\S^{[g,N]} :(P_{1},z_{1}),\ldots ,
(P_{N},z_{N})}{1,\ldots,N}{}$,
where N is the number of punctures and g the genus of the surface and the 
$(P_{i},z_{i})$:s are the punctures and local coordinates. To each local 
coordinate we 
associate a surface $\O_{i}$; $\O_{i}\subset{\bf C}\cup\{\infty\}$. The surfaces 
$\O_{i}$ are chosen such that
, when mapped to
the abstract Riemann surface, they cut out a disjoint set of regions which cover 
the Riemann
surface.   

The surface state is thus a 
tensor
${\cal T}^{(N,0)}\equiv\bigotimes^{N}{\cal H}^{*}$, where $\cal H^{*}$ is the 
dual of  
the state space $\cal H$ of the local CFT. The correlation functions on the 
Riemann 
surface $\S$ is then accomplish by saturating the tensor as:
\eq{\big{<}\F_{1}\ldots\F_{N}\big{>}_{\S}=\bra{\S^{[g,N]} :(P_{1},z_{1}),\ldots ,
(P_{N},z_{N})}
{1,\ldots,N}{}\ket{\F_{1}}{1}{}\ldots\ket{\F_{N}}{N}{}\eqn{OF1}}

A surface state of particular importance is the reflector,
\eq{\bra{R}{12}{}\equiv\bra{{\bf P}^{1}:({z_{1}},z_{2}=\frac{1}{z_{1}})}{}{}
\eqn{reflector}}
where ${\bf P}^{1}\equiv{\bf C}\cup\{\infty\}$, {\it i.e.} $\bra{R}{12}{}$ is a sphere 
with two
punctures. It is used in
the sewing of two surface states
\eq{\bra{\S\infty\S'}{1...N,N+1...N+M}{}=
\bra{\S\cup(P_{i},z^{(i)})} {1...N,i}{}
\bra{\S'\cup(P_{j},z^{(j)})} {N+1...N+M,j}{}\ket{R}{ij}{}\eqn{sewing}}
and it reflects the external states
\eq{\bra{\F_{i}}{1}{}\ket{R}{12}{}=\ket{\F_{i}}{2}{}\eqn{reflection}} 

One of the major properties of the surface states are the overlap conditions:     

\eq{\bra{\S^{[g,N]} :(P_{1},z_{1}),\ldots ,(P_{N},z_{N})}{1,\ldots,N}{}(\F^{(i)}(z_{i})
dz_{i}^{\D_{\F}}-\F^{(j)}(z_{j})dz_{j}^{\D_{\F}})=0\eqn{overlap}}
where $\D_{\F}$ is the conformal weight of the field $\F$. The overlap conditions imply
that the surface states are indeed realizations of the abstract Riemann surfaces. 

We end this section with the definition of the Virasoro generators for a $c=0$ CFT:
\eq{\bra{R}{12}{}({\bf 1}+\e_{n} L^{(1)}_{n}+\bar{\e}_{n} \bar{L}^{(1)}_{n})=
\bra{{\bf P}^{1}:
({z=z_{1}+\e_{n} z_{1}^{n}},z_{2}=\frac{1}{z})}{}{}\equiv \bra{P_{n}}{}{}
\eqn{Virasoro}}
where $z$ is the universal coordinate of the sphere.
From this definition it's easy to derive the $c=0$ Virasoro algebra.

\section{Analytical Regularization}\label{analyticalregularization}

We consider a $c=0$ CFT in the operator formulation.
Let $\ket{\F_{i}}{}{}=\hat{\F}_{i}(0,0)\ket{)}{}{}$ be a basis for ${\cal H}$, the 
state space of the CFT,
and $\ket{\F_{\m}}{}{}=\hat{\F}_{\m}(0,0)\ket{0}{}{}$ be a basis for
the physical subspace ${\cal H}^{phys}$ of ${\cal H}$.

We suppose that the CFT allows analytical regularization of surface integrals
of the correlation functions.
This regularization method is essentially the same as the one used for
calculations of amplitudes in ordinary string theory with a continuous 
$\hat{L}_{0}$
spectrum due to a continuous particle momentum in the uncompactified directions.
In \cite{cvgs} we expressed the analytical regularization method as
\eq{\int_{{\bf C}}d^{2}z\hat{\F}_{\m}(z,\bar{z})	=0\eqn{regularization}}
To see what is really going on in \refbr{regularization} let's consider a general 
vertex operator $\hat{\F}_{\m}(z,\bar{z})$ of weight $(1,1)$ carrying left and right 
momentum 
$k$. It can be given a mode expansion as
\eq{\hat{\F}_{\m}(z,\bar{z}) = \sum_{m,n\in{Z}}\hat{\F}_{m,n,\m}|z|^{2\g}
z^{-1-m}\barz^{-1-n}\eqn{areg11}}
where $\g$ is the operator valued shift $(k\cdot\hat{p})\over{4}$.
When integrating $\hat{\F}_{\m}(z,\bar{z})$ over a disc with radius $R$ we get

\eqs{-\frac{1}{\p}\int_{|z|\leq R}d^{2}z\hat{\F}_{\m}(z,\bar{z})&=&
-\frac{1}{\p}\int_{|z|\leq R}d^{2}z\sum_{m,n\in{Z}}\hat{\F}_{m,n,\m}|z|^{2\g}
z^{-1-m}\barz^{-1-n}\lb 
&=&-\frac{1}{\p}\int_{0}^{R}rdr\int_{0}^{2\p}d\th
\sum_{m,n\in{Z}}\hat{\F}_{m,n,\m}r^{2\g-2-m-n}e^{i(n-m)\th}\lb
&=&-\sum_{n\in{Z}}\frac{R^{2(\g -n)}}{\g-n}\hat{\F}_{n,n,\m}\eqn{areg12}}
after having used the analytical regularization
\eq{\int_{0}^{1}dx x^{\a} = \frac{1}{1+\a},\quad\a\neq-1\eqn{areg13}}
in the radial integration. It's now easy to see, using
 
\eq{\int_{0}^{\infty}dx x^{\a}=\int_{0}^{1}+\int_{1}^{\infty}dx x^{\a}
=\int_{0}^{1}dx(x^{\a}+x^{-2-\a})=\frac{1}{1+\a}-\frac{1}{1+\a}=0
\eqn{areg14}}
that an operator $\hat{\F}_{\m}(z,\bar{z})$ having singularities only at zero 
and infinity obeys \refbr{regularization}.
 
When we expand \refbr{regularization}
in the basis of ${\cal H}$ we sandwich
$\int_{{\bf C}}d^{2}z\hat{\F}_{\m}(z,\bar{z})$ between an
in-state $\ket{\F_{j}}{}{}$ at $z=0$ and an out-state $\bra{\F_{i}}{}{}$ 
at $z=\infty$. This matrix element is given by the radial ordered integral
\eq{\int_{{\bf C}}\bra{\F_{i}}{}{}\hat{\F}_{\m}
\ket{\F_{j}}{}{}=0\eqn{areg3}} 
More generally, if we saturate $\int_{{\bf C}}d^{2}z\hat{\F}_{\m}(z,\bar{z})$ 
with an in state $\hat{\F}_{j}(w,\bar{w})
\ket{0}{}{}$ at $w$ and
an out-state $\bra{0}{}{}\hat{\F}_{i}(u,\bar{u})$ at $u$ then 
we get $\int_{{\bf C}}\vev{{\cal R}(\hat{\F}_{i}(u,\bar{u})
\hat{\F}_{\m}(z,\bar{z})\hat{\F}_{j}(w,\bar{w}))}{}$, which can be brought to the
canonical form \refbr{areg3} by a M\"{o}bius transformation.
It's important to notice that \refbr{regularization} is not valid when 
the integral appears radially ordered together with more than
two other conformal fields.
In this case one has to divide the integration region 
into well-ordered subregions, each of which
does not vanish.

As a consequence of \refbr{regularization} and the operator product expansion rules
we can in fact summarize the analytical regularization method in
the more general rule:
\eq{\int_{{\bf C}^{N}} {\cal R}(\hat{\F}_{\m_{1}}\cdots\hat{\F}_{\m_{N}})=0
\eqn{generalregularization}}
Here and in the sequel we use the shorthand notation:
\eqs{&{}&\int_{R_{1}\times R_{2}\times\cdots\times R_{n}}{\cal R}(\hat{\F}_{i_{1}}
\hat{\F}_{i_{2}}\cdots\hat{\F}_{i_{n}})\equiv\\
&{}&\int_{z_{1}\in R_{1},\;z_{2}\in R_{2},\; ....\; ,z_{n}\in R_{n}}
d^{2}z_{1}d^{2}z_{2}\cdots d^{2}z_{n}{\cal R}(\hat{\F}_{i_{1}}(z_{1},\bar{z}_{1})
\hat{\F}_{i_{2}}(z_{2},\bar{z}_{2})\cdots\hat{\F}_{i_{n}}(z_{n},\bar{z}_{n}))}
where $R_{i}$ are regions in ${\bf C}$.

Another way of explaining \refbr{generalregularization} is to point out that after 
$N-1$ 
integrations over ${\bf C}$ of ${\cal R}(\hat{\F}_{\m_{1}}\cdots\hat{\F}_{\m_{N}})$
we obtain the generator of $N+2$ point functions on the sphere, so that in the
final integration over ${\bf C}$ the integrand is a number, and hence
the integral vanishes as a special case of \refbr{regularization}.

\section{Marginal Deformations of Conformal Field Theory}\label{marginaldeformation}

It is well known that
the Physical space ${\cal H}^{phys}$ is (isomorphic to) the tangent space 
of a space of deformed conformal field theories. 
This space can be parameterized by local coordinates $t^{\m}$, 
such that the action of the derivative $\*_{\m}$ 
on a correlation function of some  
conformal operators on a Riemann surface $\S$
is given by the insertion of the $\S$-integral of
the marginal operator $\hat{\F}_{\m}$ into the correlator:
\eq{\*_{\m}\vev{{\cal R}(\hat{\F}_{i_{1}}\cdots \hat{\F}_{i_{N}})}{\S}(t)=
\vev{{\cal R}(\int_{\S}\hat{\F}_{\m}\hat{\F}_{i_{1}}\cdots \hat{\F}_{i_{N}})}{\S}(t)
\eqn{deformedcorr}}
This so called marginal deformation can be shown to generate new CFT's.
Equation \refbr{deformedcorr} will be derived in section \ref{definitionconnection}
from the definition of a connection.

There is a question of regularization in \refbr{deformedcorr}, since the surface 
integral 
contains singularities which can only be defined using some regularization 
method. 
If the formal expression that we regularize here
is to be the action of a local commuting basis of vector fields $\*_{\m}$, on a 
local function $\vev{{\cal R}(\cdots)}{\S}(t)$,
then the ``arbitrariness'' in the  ``subtractions'' involved in
the regularization procedure have to be carefully absorbed by the vector 
fields $\*_{\m}$ such that no ``anomalies''
occur in the commutator $[\*_{\m},\*_{\n}]$.
By virtue of the analytical 
regularization method the vector fields $\*_{\m}$ are indeed commuting
\footnote{We assume that the CFT is local
so that correlation functions have trivial monodromy}:
\eq{[\*_{\m},\*_{\n}]=0\eqn{derivativescommute}}
This of course means that we can integrate the vector fields $\*_{\m}$ 
to obtain the usual expression for the deformed correlation functions:
\eq{\vev{{\cal R}(\hat{\F}_{i_{1}}\cdots \hat{\F}_{i_{N}})}{\S}(t)=
\vev{{\cal R}(\exp(t^{\m}\int_{\S}\hat{\F}_{\m})
\hat{\F}_{i_{1}}\cdots \hat{\F}_{i_{N}})}{\S}(0)\eqn{deformedcorrelator}}
In equation \refbr{deformedcorrelator} the only $t$ dependence lies in the exponent. 
This means that we have made a choice of basis $\ket{\F_{i}}{}{}$ which 
by assumption is $t$ independent . Of course we can always make a different choice 
of basis $\ket{\F'_{i}}{}{}=\L_{i}^{j}(t)\ket{\F_{i}}{}{}$. In this $t$ dependent basis
the derivative of the correlation function would pick up an extra factor due to the 
derivative of $\L_{i}^{j}(t)$. This might seem to be a nuisance but actually causes no
harm since the crucial condition, the commutation of 
vector fields $\*_{\m}$, is also satisfied in this basis.

In other circumstances where analytical regularization isn't applicable it may happen
that anomalies occur in the commutator $[\*_{\m},\*_{\n}]$. The ordinary 
derivative $\*_{\m}$ then has to be reinterpreted as the 
covariant derivative ${\cal D}_{\m}$, and the anomaly as curvature. This 
interpretation is all right since the correlation function are just ordinary 
tensor coefficient carrying indices. We will come back to this point in section
\refbr{curvature} where we determine the covariant derivative for a set of 
analytically regularized connections.

\section{Abstract Formulation of CFT Deformations}

Since we are interested in deformations of conformal field theories, we
parameterize the space of CFT $\cal C$ by coordinates 
$t^{\m}$, as stated above.
At each point $t^{\m}$ we then label the state space {\it i.e.} the Hilbert space ${\cal H}$
by $\HH_{t}$. In fact $\HH$ is a vector fiber over the theory space ${\cal C}$.

In order to compare theories at different points in theory space,
we need to introduce a parallel transport,
\eq{\s_{t_{0}\la t}^{*}:\HH_{t}\ra \HH_{t_{0}}\eqn{af1}}
The parallel transport generalizes to an arbitrary tensor $T\in {\cal T}^{n,m}
\equiv\bigotimes^{n}\HH^{*}\bigotimes^{m}\HH$ , by acting on each 
Hilbert space separately.

The parallel transport is generated by a connection $\na $. If we 
let $\ket{\F_{i}}{}{}:\CC\ra H$ be a section then the connection is defined 
through
\eq{\s_{t\la t+dt}^{*}\ket{\F_{i}}{}{t+dt}=\ket{\F_{i}}{}{t}+
dt^{\m}\na_{\m}^{t}\ket{\F_{i}}{}{t}\eqn{af2}}
The generalization to arbitrary tensors implies that $\na_{\m}$ obeys the Leibniz'
rule.

In order for the parallel transport to be well defined on $\cal C$ it 
has to commute with 
the sewing operation {\it i.e.},
\eqs{\s_{t_{0}\la t}^{*}(\bra{\S\infty\S'}{1...N,N+1...N+M}{\quad\quad\quad\quad
\quad\;\; t})=
\s_{t_{0}\la t}^{*}(\bra{\S\cup(P_{i},z^{(i)})} {1...N,i}{\quad\quad t})
\s_{t_{0}\la t}^{*}(\bra{\S'\cup(P_{j},z^{(j)})} {N+1...N+M,j}{\quad\quad\quad
\quad\; t})\lb
\s_{t_{0}\la t}^{*}(\ket{R}{ij}{t})\eqn{parallelsewing}}
since otherwise the parallel transport would not 
preserve the axioms of a CFT.

By the introduction of a connection we can generate infinitesimal 
transformation. However we would like to make comparisons of different 
theories lying at finite distances apart. We thus need to generate a finite
parallel transport and for this we split the interval $[t_{0},t_{N}]$ into N 
pieces.
The parallel transport on an arbitrary tensor $T$ from $t_{N}$ to $t_{0}$
then becomes,
\eq{\s_{t_{0}\la t_{N}}^{*}(T^{t_{N}})=\P_{i=1}^{N}
\s_{t_{i-1}\la t_{i}}^{*}(T^{t_{N}})\eqn{af4}}

We now make the assumption that the action of the connection on 
an arbitrary tensor can be
written as the an operator $\hat{X}$ acting on the tensor. This operator  
is then assumed to be parallel transported as
\eq{\s_{t\la t+dt}^{*}\hat{X}^{t+dt}=\hat{X}^{t}+dt^{\m}\na_{\m}^{t}\hat{X}^{t}
\eqn{af5}}
Thus  
\eqs{\s_{t_{0}\la t_{N}}^{*}(T^{t_{N}})&=&\P_{i=1}^{(N-1)}
\s_{t_{(i-1)}\la t_{i}}^{*}\s_{t_{(N-1)}\la t_{N}}^{*}(T^{t_{N}})\lb
&=&\P_{i=1}^{N-2}\s_{t_{(i-1)}\la t_{i}}^{*}\s_{t_{(N-2)}\la t_{(N-1)}}^{*}
(1+\D t_{(N-1)}^{\m}\na_{\m}^{t_(n-1)}){T}^{t_{(N-1)}}\lb
&=&\P_{i=1}^{N-2}\s_{t_{(i-1)}\la t_{i}}^{*}(1+\D t_{N-2}^{\n}
\na_{\n}^{t_{(N-2)}})(1+
\D t_{N-1}^{\m}\na_{\m}^{t_{(N-2)}})T^{t_{(N-2)}}\eqn{af6}}
from which we conclude that
\eq{\s_{t_{0}\la t_{N}}^{*}(T^{t_{N}})={\cal P}exp(\int_{C'}dt^{\m}
\na_{\m}^{t_{0}})T^{t_{0}}\eqn{finitetrans}}
In general the path ordered expression \refbr{finitetrans} depends on the path $C'$
taken. The exception is of course when the connection is flat. In the next section we
will define explicit connections, which in section \ref{curvature} 
 will be shown to be flat and
thus path independent.   

In the sequel we will suppress the index $t$ unless it's needed for completeness.

\section{The Natural Connections $\nabla^{(R)}$ and $\d$}\label{definitionconnection}

We are now ready to define our connections which will generate CFT 
deformations.
The connection $\nabla^{t(R)}$ is given by:
\eqs{\nabla_{X}^{t(R)}\bra{\S}{1\ldots N}{\quad t}&=&\bra{\S}{1\ldots N}
{\quad t}\sum_{i=1}^{N}
X^{\m}\int_{\O_{i}\setminus D^{(R)}}\hat{\F}_{\m}^{(i)t}\eqn{nablaRsurfstate}\\
\nabla^{t(R)}_{X}\ket{\F_{i}}{}{t}&=&
X^{\m}\int_{D^{(R)}}\hat{\F}_{\m}^{t}\ket{\F_{i}}{}{t}
\eqn{nablaRextstate}}
We see that the action of $\nabla_{X}^{t(R)}$ on a surface state implies the insertion
of a marginal operator integrated over the surface $\S$ with the discs $D^{(R)}$ 
cut out around each puncture. However it's important to realize that the subtraction
of the discs does not serve as a regularization, since the discs are being 
integrated over when the connection is acting on the external states. The 
integrals should instead be evaluated using the analytical regularization defined 
in section 
\ref{analyticalregularization}. The choice of radius $R$ will be shown to be 
unimportant, in the sense that connections with different radius will generate 
transformations differing only by inner transformations. The limit $R\ra 0$
can however not be taken trivially, as will be seen below. Therefore the following 
definition of the connection $\d$ is in order
\eqs{\d_{X}^{t}\;\bra{\S}{1...N}{\quad t}&=&\bra{\S}{1...N}{\quad t}\sum_{i=1}^{N}
X^{\m}\int_{\O_{i}}\hat{\F}_{\m}^{(i),t}\eqn{deltasurfstate}\\
\d_{\m}^{t}\ket{\F_{i}}{}{t}&=&0\eqn{deltaextstate}}
This is just the connection considered in \cite{cvgs}.

An implication of \refbr{nablaRsurfstate}, \refbr{nablaRextstate}, 
\refbr{deltasurfstate} and \refbr{deltaextstate} is that the dual
state space ${\cal H}^{*}$ and the correlation functions transform as:   
\eqs{\nabla_{X}^{t(R)}\;\bra{\F_{i}}{}{t}&=&\bra{\F_{i}}{}{t}
(-X^{\m}\int_{D^{(R)}}\hat{\F}_{\m}^{t})=\bra{\F_{i}}{}{t}
(X^{\m}\int_{{\bf C}\setminus D^{(R)}}\hat{\F}_{\m}^{t})\eqn{nablaRlextstate}\\
\d_{\m}^{t}\bra{\F_{i}}{}{t}&=&0\eqn{deltalextstate}}\\
and
\eqs{\nabla_{X}^{t(R)}\bra{\S}{1\ldots N}{\quad t}\ket{\F_{i_{1}}}{1}{t}\ldots
\ket{\F_{i_{N}}}{N}{t}&=&\bra{\S}{1\ldots N}{\quad t}\sum_{i=1}^{N}
X^{\m}\int_{\O_{i}}\hat{\F}_{\m}^{(i)t}\ket{\F_{i_{1}}}{1}{t}\ldots
\ket{\F_{i_{N}}}{N}{t}=\lb
&=&\vev{{\cal R}(X^{\m}\int_{\S}\hat{\F}_{\m}
\hat{\F}_{i_{1}}\ldots\hat{\F}_{i_{N}})}{\S}\eqn{nablaRcorr}\\
\delta_{X}^{t}\bra{\S}{1\ldots N}{\quad t}\ket{\F_{i_{1}}}{1}{t}\ldots
\ket{\F_{i_{N}}}{N}{t}&=&\vev{{\cal R}(X^{\m}\int_{\S}\hat{\F}_{\m}
\hat{\F}_{i_{1}}\ldots\hat{\F}_{i_{N}})}{\S}\eqn{deltacorr}}

Here we arrive at the promised result \refbr{deformedcorr} which puts the deformation 
of the correlation function on a nice geometrical foundation.

What is so special about the connections defined above, and isn't the tensorial 
property of say \refbr{nablaRextstate} incorrect? It looks as if the left hand side of
\refbr{nablaRextstate} transforms inhomogeneously while the right hand side transforms
homogeneously under a change of basis. This is true, but equation \refbr{nablaRextstate}
should not be and as we have just mentioned can not be interpreted covariantly. 
It's only valid with respect to the 
particular basis chosen. In a generic 
basis the connection coefficients will contain terms that not only contain the 
insertion  of marginal operators but also terms depending on the transformation matrix 
between the new basis and the particular basis chosen.
This is part of the answer to our first question. The connections above all belong to
the class of natural connections {\it i.e.} connections whose structure coefficients only 
depend on data from the CTF itself, such as the conformal weights and the 
operator algebra (still only within the particular basis chosen of course). 
In addition these connections all satisfy the sewing condition
\refbr{parallelsewing} as will be shown in section \ref{sewingandconformal}. This is
of course a necessary condition for the connection to be well defined on a space of 
CFT's. Being natural the CFT itself also gives us a chance of analyzing the
connections 
without any additional data. Therefore let's proceed with this analysis.  

Naively we might think that the limit $lim_{R\ra 0}\na_{\m}^{(R)}$ is equal to 
$\d_{\m}$. 
However from what follows we will see that this is not true. Consider:
\eqs{\na_{\m}^{(R)}\ket{\F_{i}}{}{}
&=&\int_{|z|\leq R}d^{2}z\hat{\F}_{\m}(z,\bar z)
\hat{\F}_{i}(0,0)\ket{0}{}{}\lb
&=&\sum_{j}C_{\m i}^{\quad j}\int_{|z|\leq R}d^{2}z\;z^{(-1-\D_{i}+\D_{j})}
\bar{z}^{(-1-\bar{\D}_{i}+\bar{\D}_{j})}\ket{\F_{j}}{}{}\lb 
&=&\sum_{j}2\p C_{\m i}^{\quad j}\frac{\d_{\D_{j}-\bar{\D}_{j},\D_{i}-
\bar{\D}_{i}}}{\D_{j}+
\bar{\D}_{j}-\D_{i}-\bar{\D}_{i}}R^{{\D}_{j}+\bar{\D}_{j}-\D_{i}-\bar{\D}_{i}}
\ket{\F_{j}}{}{}\lb
&=&\sum_{j}2\p C_{\m i}^{\quad j}\frac{\d_{s_{i},s_{j}}}{\g_{j}-\g_{i}}R^{\g_{j}-\g_{i}}
\ket{\F_{j}}{}{}\eqn{derivconncoef}}
where $s_{i}=\D_{i}-\bar{\D}_{i}$ and $\g_{i}=\D_{i}+\bar{\D}_{i}$ and the analytical
regularization was being used.
The connections coefficients are defined as 
\eq{\na_{\m}^{(R)}\ket{\F_{i}}{}{}
=\sum_{j}\G_{\m i}^{(R)j}\ket{\F_{j}}{}{}\eqn{defconncoef}}
which of course implies that 
\eq{\G_{\m i}^{(R)j}=2\p C_{\m i}^{\quad j}
\frac{\d_{s_{i},s_{j}}}{\g_{j}-
\g_{i}}R^{\g_{j}-\g_{i}}\eqn{conncoef}}
Hence we arrive at the important result that the residue 
\eq{Res_{\g_{i}=\g_{j}}\G_{\m i}^{(R)j}=2\p C_{\m i}^{\quad j} 
\d_{s_{i},s_{j}}\eqn{resconncoef}}
is independent of $R$. The conclusion of this is that, in the analytically regularized
sense $\G_{\m i}^{(R)j}$ is independent of $R$, and therefore the limit $lim_{R\ra 0}
\na_{\m}^{(R)}$ of $\na_{\m}^{(R)}$ is not equivalent to $\d_{\m}$. Moreover the 
operator
\eqs{\hat{L}_{X}^{(R,R')}&=&\na_{X}^{(R)}-\na_{X}^{(R')}\lb
\hat{L}_{\m i}^{(R,R')j}&=&2\p C_{\m i}^{\quad j}\frac{\d_{s_{i},s_{j}}}{\g_{j}-
\g_{i}}(R^{\g_{j}-\g_{i}}-R'^{\g_{j}-\g_{i}})\eqn{diffnablaRR'}}
has zero residue at $\g_{i}=\g_{j}$ and thus generate inner transformations.
On the other hand the 
operator 
\eqs{\hat{L}_{X}^{(R)}&=&\na_{X}^{(R)}-\d_{X}\lb
\hat{L}_{\m i}^{(R)j}&=&2\p C_{\m i}^{\quad j}\frac{\d_{s_{i},s_{j}}}{\g_{j}-
\g_{i}}R^{\g_{j}-\g_{i}}\eqn{diffnablaRdelta}} 
has non-zero residue at $\g_{i}=\g_{j}$ and thus generate outer transformations.

As a particular case of \refbr{nablaRsurfstate} and \refbr{deltasurfstate} we find
\eq{\nabla^{t(R)}_{X}\;\bra{R}{12}{\;\; t}=
\bra{R}{12}{\;\; t}X^{\m}\int_{R\leq |z^{(1)}| \leq {1 \/ R}}
\hat{\F}_{\m}^{(1),t}\eqn{nablaRreflector}}
and
\eq{\d^{t}_{X}\;\bra{R}{12}{\;\; t}=
\bra{R}{12}{\;\; t}X^{\m}\int_{{\bf C}}
\hat{\F}_{\m}^{(1)t}\eqn{deltaRreflector}}
where we have used the overlap condition \refbr{overlap} and the fact that 
$\{z_{1}(z_{2});z_{2}\in \O_{2}\setminus D^{R}\}=\{1\leq z_{1}\leq\frac{1}{R}\}$.
In a later section we will show that $\bra{R}{12}{\;\; t}X^{\m}\int_{{\bf C}}
\hat{\F}_{\m}^{t}=0$ and
thus we see that $\na^{(1)}$ and $\delta$ are metric compatible. For the same reason as
above however the limit $lim_{R\ra 0} \na^{R}$ is not metric compatible.

We would now like to use the connections $\na^{R}$ and $\d$ to determine 
explicit expressions for the finite transformation \refbr{finitetrans} and 
check the sewing condition \refbr{parallelsewing}. However before we are ready to do
so we will have to go through a number of derivations.

\section{Infinitesimal Transformations of the Virasoro Generators $\hat{L}_{n}$
and the Vertex Operators $\hat{\F}_{i}$.}

The first thing we have to derive is the action of the connection on 
the Virasoro generators. We will do this rather explicit in order for the reader
to see how the analytical regularization works. If we use equation \refbr{Virasoro} 
we find:
\eqs{&{}&\e\bra{R}{12}{\;\; t}(\nabla^{t(R)}_{\m}\hat{L}^{(1)t}_{n})=
\bra{R}{12}{\;\; t}(\nabla^{t(R)}_{\m}({\bf{1}}+\e\hat{L}^{(1)t}_{n}))\lb
&=&\nabla^{t(R)}_{\m}(\bra{R}{12}{\;\; t}({\bf{1}}+\e\hat{L}^{(1)t}_{n}))-
(\nabla^{t(R)}_{\m}\bra{R}{12}{\;\; t})({\bf{1}}+\e\hat{L}_{n}^{(1)t})=\lb
&=&\bra{R}{12}{\;\; t}({\bf{1}}+\e\hat{L}^{(1)t}_{n})(\int_{\O_{1}^{'}
\setminus{D^{(R)}}}
\hat{\F}_{\m}^{(1)t}+\int_{\O_{2}\setminus{D^{(R)}}}\hat{\F}_{\m}^{(2)t})-\lb
&-&\bra{R}{12}{\;\; t}(\int_{\O_{1}\setminus{D^{(R)}}}
\hat{\F}_{\m}^{(1)t}+\int_{\O_{2}\setminus{D^{(R)}}}\hat{\F}_{\m}^{(2)t})
({\bf{1}}+\e\hat{L}^{(1)t}_{n})=\lb
&=&\bra{R}{12}{\;\; t}(\int_{\O_{1}^{'}}
\hat{\F}_{\m}^{(1)t}-\int_{\O_{1}}\hat{\F}_{\m}^{(1)t})+\e\bra{R}{12}{\;\; t}
[\int_{D^{R}}\hat{\F}_{\m}^{(1)t},\hat{L}^{(1)t}_{n}]+
\hat{L}^{(1)t}_{n}(\int_{\O_{1}}\hat{\F}_{\m}^{(1)t}+\int_{\O_{2}}\hat{\F}_{\m}^{(2)t})
=\lb
&=&\bra{R}{12}{\;\; t}(\int_{\O_{1}^{'}}
\hat{\F}_{\m}^{(1)t}-\int_{\O_{1}}\hat{\F}_{\m}^{(1)t})+\e\bra{R}{12}{\;\; t}
([\int_{D^{R}}\hat{\F}_{\m}^{(1)t},\hat{L}^{(1)t}_{n}]+[\hat{L}^{(1)t}_{n},
\int_{\O_{1}}\hat{\F}_{\m}^{(1)t}])=\lb
&=&\e\bra{R}{12}{\;\; t}
([\int_{D^{R}}\hat{\F}_{\m}^{(1)t},\hat{L}^{(1)t}_{n}])\eqn{derivnablavir}}

{\it i.e.}
\eq{\nabla^{t(R)}_{\m}\hat{L}_{n}^{t}=
[\int_{D^{(R)}}\hat{\F}^{(1)t}_{\m},\hat{L}^{(1)t}_{n}]=
\int_{D^{(R)}}\hat{\F}^{(1)t}_{\m}\hat{L}^{(1)t}_{n}+\hat{L}^{(1)t}_{n}
\int_{{\bf C}\setminus D^{(R)}}\hat{\F}^{(1)t}_{\m}\eqn{nablaRVir}}
Here the integration region $\O_{i}^{'}$ is $|z_{1}+\e z_{1}^{n}|<1$ while $\O_{i}$ is
$|z_{1}|<1$. Using this it is then easy to show that $(\int_{\O_{1}^{'}}
\hat{\F}_{\m}^{(1)t}-\int_{\O_{1}}\hat{\F}_{\m}^{(1)t})=-\e[\hat{L}^{(1)t}_{n},
\int_{\O_{1}}\hat{\F}_{\m}^{(1)t}]$. When deriving 
finite deformations of the Virasoro generators and showing conformal invariance it is
the last expression in \refbr{nablaRVir} that is most suitable.

We can now determine the change of the conformal weights of the conformal fields, along
an arbitrary path in theory space.
For this we use:
\eq{\hat{L}^{t+dt}_{n}=\s^{*}_{t+dt\la t}(\hat{L}^{t}_{n}+dt^{\m}\na^{t(R)}_{\m}
\hat{L}^{t}_{n})\eqn{inftVirasoro}}
\eq{\ket{\F_{i}}{}{t+dt}=\s^{*}_{t+dt\la t}(\ket{\F_{i}}{}{t}+dt^{\m}\na^{t(R)}_{\m}
\ket{\F_{i}}{}{t})\eqn{inftextstate}}
In section \ref{sewingandconformal} we will show that the finitely deformed Virasoro 
generators obey the same algebra as the undeformed operators.
The action of the Virasoro generator at $t+dt$ is then determined by:
\eqs{\hat{L}^{t+dt}_{n}\ket{\F_{i}}{}{t+dt}&=&\s^{*}_{t+dt\la t}\{(\hat{L}^{t}_{n}
+dt^{\m}\na^{t(R)}_{\m}\hat{L}^{t}_{n})(\ket{\F_{i}}{}{t}+dt^{\m}\na^{t(R)}_{\m}
\ket{\F_{i}}{}{t})\}\lb 
&=&\s^{*}_{t+dt\la t}\{(\hat{L}^{t}_{n}
+dt^{\m}[\int_{D^{(R)}}\hat{\F}^{(1)}_{\m},\hat{L}^{(1)t}_{n}])(\ket{\F_{i}}{}{t}+
dt^{\n}\int_{D^{(R)}}\hat{\F}_{\n}^{t}\ket{\F_{i}}{}{t})\}\lb
&=&\s^{*}_{t+dt\la t}\{(1+dt^{\n}\int_{D^{(R)}}\hat{\F}_{\n}^{t})\hat{L}^{t}_{n}
\ket{\F_{i}}{}{t}\}
\eqn{derivconformweight}}
Suppose now that $\ket{\F_{i}}{}{}$ is a primary field at $t^{\m}$ {\it i.e.},
 $\hat{L}^{t}_{n}
\ket{\F_{i}}{}{t}=0$ when $n\geq 1$ and $\hat{L}^{t}_{0}
\ket{\F_{i}}{}{t}=\D_{i}^{t}\ket{\F_{i}}{}{t}$, then we arrive at the conditions:
\eqs{\hat{L}^{t+dt}_{n}\ket{\F_{i}}{}{t+dt}&=&0\quad n\geq 1\lb
\hat{L}^{t+dt}_{0}\ket{\F_{i}}{}{t+dt}&=&\D_{i}^{t}\ket{\F_{i}}{}{t+dt}
\eqn{conformalweight}}
Hence the  primary fields at a point $t^{\m}$ remain primary, with the same weights, 
along any path in theory space. In particular marginal operators remain marginal. This
is of fundamental importance for the integrability of the deformed surface and external
states. If the marginal operators had not remained marginal the deformation of the 
CFT would have required a new set of marginal operators at each point in theory space
making the parallel transport extremely hard to perform.

In order to compute finite deformations we will need the expression for $\na^{(R)}$ 
acting on a vertex function. The calculation can easily be computed by
expanding the operator using 
${\bf 1}=\sum_{i}\ket{\F^{i}}{}{}\bra{\F_{i}}{}{}$. We  
get
\eqs{\nabla^{(R)}_{X}\hat{\F}_{i}(z,\bar{z})&=&X^{\m}(\int_{{\bf C}}
{\cal R}(\hat{\F}_{\m}\hat{\F}_{i}(z,\bar{z}))+[\int_{D^{(R)}}\hat{\F}_{\m},
\hat{\F}_{i}(z,\bar{z})])\eqn{nablaRop}\\
\d^{t}_{X}\hat{\F}_{i}(z,\bar{z})&=&X^{\m}\int_{{\bf C}}
{\cal R}(\hat{\F}_{\m}\hat{\F}_{i}(z,\bar{z}))\eqn{deltaop}}
We see that \refbr{deltaop} is the relation already obtained in \cite{cvgs}. 
Before we make any further comments about \refbr{nablaRop} and \refbr{deltaop}
we can, as a check of consistency, also derive \refbr{nablaRop}
starting from the expression:
\eqs{\bra{P}{123}{}\ket{\F_{j}}{3}{}&=&\bra{P}{123}{}
\hat{\F}_{j}^{(3)}(z_{3}=0,\bar{z}_{3}=0)\ket{0}{3}{}\lb
&=&\bra{P}{123}{}\ket{0}{3}{}({dz_{1}\/ dz_{3}})^{\D_{j}}
({d\bar{z}_{1}\/ d\bar{z}_{3}})^{\bar{\D}_{j}}\hat{\F}_{j}^{(1)}(z_{1}(z_{3}=0),
\bar{z}_{1}(\bar{z}_{3}=0))\eqn{consnablaRop}}
where $\bra{P}{123}{}$ is the three punctured sphere.
Making use of \refbr{conformalweight} 
{\it i.e.} $\nabla^{(R)}_{\m}\D_{j}=\*_{\m}\D_{j}=0$,
we then find:
\eqs{&{}&\bra{P'}{12}{}(\nabla^{(R)}_{\m}\hat{\F}_{j}^{(1)})(z_{1}
(z_{3}=0),\bar{z}_{1}(\bar{z}_{3}=0))=\lb
&=&({dz_{1}\/ dz_{3}})^{-\D_{j}}
({d\bar{z}_{1}\/ d\bar{z}_{3}})^{-\bar{\D}_{j}}
\nabla^{(R)}_{\m}(\bra{P}{123}{}\ket{\F_{j}}{3}{})-
(\nabla^{(R)}_{\m}\bra{P'}{12}{})\hat{\F}_{j}^{(1)}(z_{1},\bar{z}_{1})=\lb
&=&({dz_{1}\/ dz_{3}})^{-\D_{j}}
({d\bar{z}_{1}\/ d\bar{z}_{3}})^{-\bar{\D}_{j}}
\{(\nabla^{(R)}_{\m}\bra{P}{123}{})\ket{\F_{j}}{3}{}+
\bra{P}{123}{}(\nabla^{(R)}_{\m}\ket{\F_{j}}{3}{})\}-
(\nabla^{(R)}_{\m}\bra{P'}{12}{})\hat{\F}_{j}^{(1)}{i}
(z_{1},\bar{z}_{1})=\lb
&=&({dz_{1}\/ dz_{3}})^{-\D_{j}}
({d\bar{z}_{1}\/ d\bar{z}_{3}})^{-\bar{\D}_{j}}
\bra{P}{123}{}(\sum_{i=1}^{3}\int_{\O_{i}\setminus D^{(R)}}
\hat{\F}_{\m}^{(i)}+\int_{D^{(R)}}\hat{\F}_{\m}^{(3)})\ket{\F_{j}}{3}{}-\lb
&-&\bra{P'}{12}{}\sum_{i=1}^{2}\int_{\O_{i}'\setminus D^{(R)}}
\hat{\F}_{\m}^{(i)}\hat{\F}_{j}^{(1)}(z_{1},\bar{z}_{1})=\lb
&=&({dz_{1}\/ dz_{3}})^{-\D_{j}}
({d\bar{z}_{1}\/ d\bar{z}_{3}})^{-\bar{\D}_{j}}
\bra{P}{123}{}\ket{\F_{j}}{3}{}\sum_{i=1}^{2}\int_{\O_{i}'\setminus D^{(R)}}
\hat{\F}_{\m}^{(i)}-\bra{P'}{12}{}\sum_{i=1}^{2}\int_{\O_{i}'\setminus D^{(R)}}
\hat{\F}_{\m}^{(i)}\hat{\F}_{j}^{(1)}(z_{1},\bar{z}_{1})=\lb
&=&\bra{P'}{12}{}\hat{\F}_{j}^{(1)}(z_{1},\bar{z}_{1})
\sum_{i=1}^{2}\int_{\O_{i}'\setminus D^{(R)}}
\hat{\F}_{\m}^{(i)}-\bra{P'}{12}{}\sum_{i=1}^{2}\int_{\O_{i}'\setminus D^{(R)}}
\hat{\F}_{\m}^{(i)}\hat{\F}_{j}^{(1)}(z_{1},\bar{z}_{1})
\eqn{firsthalf}}
We end here. The integration regions $\O_{i},\;\;i=1,2,3$ are chosen such that they 
partition the 
three punctured sphere and such that there is precisely one puncture in each region. 
In the last lines of 
\refbr{firsthalf} we have defined $\O'_{i},\;\;i=1,2$ such that they also partition the 
two punctured sphere and such that there is only one puncture in each region. In addition
the boundary of $\O'_{i}$:s are chosen 
such that $z_{1}(z_{3}=0)\in\*\O_{1}'\equiv-\*\O_{2}'$. This implies that the first 
term in the last line
is radially ordered. Having realized this it is then easy to complete 
the calculation using the overlap
condition and the analytical regularization, finally arriving at \refbr{nablaRop}.

We now wish to make some comments 
about the role of the second term in \refbr{nablaRop}. 
The first thing to notice is that it's 
only regular at $|z|=R$. For those $z$ it actually
cancels the first radial ordered term (which is of course regular for all $z$):
\eq{\na_{\m}^{R}\hat{\F}_{i}(z,\bar{z})|_{|z|=R}=0\eqn{irregreg}}
We can therefore characterize $\na^{(R)}$ by saying that it's the natural connection 
that annihilates insertions of states at points on the circle $|z|=R$. 
Compare this to \refbr{deltaextstate}. In this sense indeed $\d=\na^{(0)}$, even though
the limit $R\rightarrow 0$ of $\na^{(R)}$ is not well-behaved. 
The matrix elements of the commutator are nevertheless in general infinite 
when taken with 
respect to the unperturbed external states. But using perturbed external states its
matrix elements precisely cancel another irregular contribution coming from the 
parallel transport of the external states.
\eqs{&&(1+dt^{\m}\na_{\m}^{(R)})\bra{\F_{i}}{}{}\hat{\F}_{j}(z,\bar{z})
\ket{\F_{k}}{}{}=\lb
&=&((1+dt^{\m}\na_{\m}^{(R)})\bra{\F_{i}}{}{})((1+dt^{\n}\na_{\n}^{(R)})\hat{\F}_{j}
(z,\bar{z}))((1+dt^{\k}\na_{\k}^{(R)})\ket{\F_{k}}{}{})=\lb
&=&\bra{\F_{i}}{}{}(1-dt^{\m}\int_{D^{(R)}}\hat{\F}_{\m})(\hat{\F}_{j}(z,\bar{z})+
dt^{\n}
\int_{\bf C}{\cal R}(\hat{\F}_{\n}\hat{\F}_{j}(z,\bar{z}))+dt^{\n}[\int_{D^{(R)}}
\hat{\F}_{\n},\hat{\F}_{j}(z,\bar{z})])\lb
&&(1+dt^{\k}\int_{D^{(R)}}\hat{\F}_{\k})\ket{\F_{k}}{}{}=\lb
&=&\bra{\F_{i}}{}{}\hat{\F}_{j}(z,\bar{z})\ket{\F_{k}}{}{}+
dt^{\m}\int_{\bf C}\bra{\F_{i}}{}{}
{\cal R}(\hat{\F}_{\m}\hat{\F}_{j}(z,\bar{z}))\ket{\F_{k}}{}{}\eqn{inftycancel}}
In the operator formulation we only use vertex operators explicitly to write down 
the covariant derivative of surface states in terms of an insertion of a surface 
integral of a marginal vertex operator equation \refbr{nablaRsurfstate}. Of course
we may rewrite this expression using an additional external state $\ket{\F_{\m}}{N=1}{}$
inserted on an additional external leg whose position is then integrated over
$\S\setminus\cup_{i}D_{i}^{(R)}$.
However the irregular term appearing in $\na_{\m}^{R}\hat{\F}_{i}(z,\bar{z})$ 
expresses the 
appearance of potentially singular contributions in higher order perturbations of the
form $\na_{\m_{1}}^{R}\ldots\na_{\m_{N}}^{R}\bra{\S}{1\ldots N}{}$, that will appear in
one form or another regardless of the choice of notation. The cancelation of these
potential singularities is a major test of the regularization method. In section 
\ref{curvature}
we will see that the cancelation works in a very simple manner in the analytically 
regularized formalism.
 
Another remark that we wish to make here is that of the consistency between the 
expression for $\na_{\m}^{(R)}\hat{T}(z)$ given in \refbr{nablaRop} and 
$\na_{\m}^{(R)}\hat{L}_{n}$ given in \refbr{nablaRVir}. If we
insert $\hat{T}(z)$ into \refbr{nablaRop} the first term becomes 
$\frac{\bar{z}}{z}\hat{\F}_{\m}(z,\bar{z})$. Consistency follows if we 
interpret $\hat{L}_{n}$ as the modes of $\hat{T}(z,\bar{z})$ at $z=0$ and if we take 
$z\ra 0$ before we expand the radially ordered integral in the expression for 
$\na_{\m}\hat{T}(z,\bar{z})$, so that this integral vanishes.

\section{Curvature, Torsion and Finite Deformations}\label{curvature}

In order to determine the curvature, torsion and finite deformation we have to consider
higher orders of the connection. This will be done in the first part of this section 
requiring some rather lengthy calculations. 
\subsection{Higher orders of the connection}
When we apply the connection repeatedly using
\refbr{nablaRsurfstate}, \refbr{nablaRextstate} and
\refbr{nablaRop} we get
products of integrals of marginal field operators. 
Schematically we can write $\nabla \bra{\S}{}{}=
\bra{\S}{}{}\int \hat{\F}$ and then $\nabla\nabla\bra{\S}{}{}=
\bra{\S}{}{}(\int \hat{\F}\int \hat{\F}+\nabla\int\hat{\F})$.
In order for the connection to generate regular 
finite deformations, the expressions of this type
must be in radial order. This requires cancelation between different contributions 
as mentioned above. To really see how these cancelation come about and to see how 
the analytical regularization take cares of all the potential problems we present 
complete derivations of the higher order terms both for the external and surface state.

For an external state $\ket{\F_{i}}{}{}$ the second order term becomes: 
\eqs{&{}&\nabla^{(R)}_{\m}\nabla^{(R)}_{\n}\ket{\F_{i}}{}{}=
\nabla^{(R)}_{\m}\int_{D^{(R)}}\hat{\F}_{\n}\ket{\F_{i}}{}{}=\lb
&=&(\int_{{\bf C}\times D^{(R)}}{\cal R}(\hat{\F}_{\m}\hat{\F}_{\n})+
[\int_{D^{(R)}}\hat{\F}_{\m},\int_{D^{(R)}}\hat{\F}_{\n}]+
\int_{D^{(R)}}\hat{\F}_{\n}\int_{D^{(R)}}\hat{\F}_{\m})\ket{\F_{i}}{}{}=\lb
&=&(\int_{D^{(R)}\times D^{(R)}}{\cal R}(\hat{\F}_{\m}\hat{\F}_{n})+
(\int_{\tilde{D}^{(R)}}\hat{\F}_{\m}+
\int_{D^{(R)}}\hat{\F}_{\m})\int_{D^{(R)}}\hat{\F}_{\n})\ket{\F_{i}}{}{}=\lb
&=&\int_{D^{(R)}\times D^{(R)}}{\cal R}(\hat{\F}_{\m}\hat{\F}_{n})
\ket{\F_{i}}{}{}
\eqn{nablasquared}}
Here we have used the short hand notation $\tilde{D}^{(R)}\equiv {\bf C}\setminus 
D^{(R)}$.

Inspired by the expression of the second ordered term we now try to prove by induction 
that
\eq{(\prod_{i=1}^{M}\nabla^{(R)}_{\m_{i}})\ket{\F_{i}}{}{}=
\int_{(D^{(R)})^M}{\cal R}(\prod_{i=1}^{M}\hat{\F}_{\m_{i}})\ket{\F_{i}}{}{}
\eqn{nablan}}
Using \refbr{nablan} we therefore compute
\eqs{&{}&\nabla_{\m}^{(R)}(\prod_{i=1}^{M}\nabla^{(R)}_{\m_{i}})
\ket{\F_{i}}{}{}=\lb
&=&(\int_{{\bf C}\times(D^{(R)})^M}
{\cal R}(\hat{\F}_{\m}\prod_{i=1}^{M}\hat{\F}_{\m_{i}})+
[\int_{D^{(R)}}\hat{\F}_{\m},\int_{(D^{(R)})^M}
{\cal R}(\prod_{i=1}^{M}\hat{\F}_{\m_{i}})]+\int_{(D^{(R)})^n}
{\cal R}(\prod_{i=1}^{M}\hat{\F}_{\m_{i}})\int_{D^{(R)}}\hat{\F}_{\m})
\ket{\F_{i}}{}{}\lb
&=&(\int_{(D^{(R)})^{M+1}}{\cal R}(\hat{\F}_{\m}
\prod_{i=1}^{M}\hat{\F}_{\m_{i}})+
(\int_{\tilde{D}^{(R)}}\hat{\F}_{\m}+
\int_{D^{(R)}}\hat{\F}_{\m})
\int_{(D^{(R)})^{M}}{\cal R}(\prod_{i=1}^{M}\hat{\F}_{\m_{i}}))
\ket{\F_{i}}{}{}=\lb
&=&(\int_{(D^{(R)})^{M+1}}{\cal R}(\hat{\F}_{\m}
\prod_{i=1}^{M}\hat{\F}_{\m_{i}})+
\int_{{\bf C}}\hat{\F}_{\m}
\int_{(D^{(R)})^{M}}{\cal R}(\prod_{i=1}^{M}\hat{\F}_{\m_{i}}))
\ket{\F_{i}}{}=\lb
&=&\int_{(D^{(R)})^{M+1}}{\cal R}
(\hat{\F}_{\m}\prod_{i=1}^{M}\hat{\F}_{\m_{i}})
\ket{\F_{i}(t)}{}{}}
which thus completes the proof of \refbr{nablan}.

For a surface state $\bra{\S}{1...N}{}$ the second order term becomes:
\eqs{&{}&\nabla^{(R)}_{\m}\nabla^{(R)}_{\n}\bra{\S}{1...N}{}=\lb
&=&\bra{\S}{1...N}{}(\sum_{i,j=1}^{N}\int_{\O_{i}\setminus D^{(R)}}
\hat{\F}^{(i)}_{\m}\int_{\O_{j}\setminus D^{(R)}}
\hat{\F}^{(j)}_{\n}+\sum_{i=1}^{N}\int_{{\bf C}\times \O_{i}\setminus D^{(R)}}
{\cal R}(\hat{\F}^{(i)}_{\m}\hat{\F}^{(i)}_{\n})+\lb
&+&\sum_{i=1}^{N}[\int_{D^{(R)}}\hat{\F}_{\m}^{(i)},\int_{\O_{i}
\setminus D^{(R)}}\hat{\F}_{\n}^{(i)}])}
In the double sum the terms for $i\neq j$ are radial ordered, while for
$i=j$ we get the term:
\eqs{&{}&\int_{\O_{i}\setminus D^{(R)}}
\hat{\F}^{(i)}_{\m}\int_{\O_{i}\setminus D^{(R)}}
\hat{\F}^{(i)}_{\n}+\int_{{\bf C}\times \O_{i}\setminus D^{(R)}}
{\cal R}(\hat{\F}^{(i)}_{\m}\hat{\F}^{(i)}_{\n})+
[\int_{D^{(R)}}\hat{\F}_{\m}^{(i)},\int_{\O_{i}
\setminus D^{(R)}}\hat{\F}_{\n}^{(i)}]=\lb
&=&\int_{\O_{i}\setminus D^{(R)}}
\hat{\F}^{(i)}_{\m}\int_{\O_{i}\setminus D^{(R)}}
\hat{\F}^{(i)}_{\n}+\int_{\tilde{\O}_{i}}
\hat{\F}^{(i)}_{\m}\int_{\O_{i}\setminus D^{(R)}}
\hat{\F}^{(i)}_{\n}+\lb
&+&\int_{(\O_{i}\setminus D^{(R)})^{2}}
{\cal R}(\hat{\F}^{(i)}_{\m}\hat{\F}^{(i)}_{\n})+
\int_{\O_{i}\setminus D^{(R)}}
\hat{\F}^{(i)}_{\n}\int_{D^{(R)}}
\hat{\F}^{(i)}_{\m}+
[\int_{D^{(R)}}\hat{\F}_{\m}^{(i)},\int_{\O_{i}
\setminus D^{(R)}}\hat{\F}_{\n}^{(i)}]=\lb
&=&(\int_{\tilde{\O}_{i}}\hat{\F}_{\m}+\int_{\O_{i}\setminus D^{(R)}}
\hat{\F}_{\m}+\int_{D^{(R)}}\hat{\F}_{\m})\int_{\O_{i}\setminus D^{(R)}}
\hat{\F}_{\n}+
\int_{(\O_{i}\setminus D^{(R)})^{2}}
{\cal R}(\hat{\F}^{(i)}_{\m}\hat{\F}^{(i)}_{\n})=\lb
&=&\int_{(\O_{i}\setminus D^{(R)})^{2}}
{\cal R}(\hat{\F}^{(i)}_{\m}\hat{\F}^{(i)}_{\n})}
Let us prove by induction that
\eq{(\prod_{k=1}^{M}\nabla_{\m_{k}}^{(R)})\bra{\S}{1...N}{}=
\bra{\S}{1...N}{}\sum_{i_{1},...,i_{M}=1}^{N}
\int_{\O_{i_{1}}\setminus D^{(R)}\times \cdots\O_{i_{M}}\setminus
D^{(R)}}{\cal R}(\prod_{k=1}^{M}\hat{\F}_{\m_{k}}^{(i_{k})})\eqn{nablansurf}}
Using \refbr{nablansurf} we therefore compute
\eqs{&{}&\nabla^{(R)}_{\m}(\prod_{k=1}^{M}\nabla_{\m_{k}})
\bra{\S}{1...N}{}=\lb
&=&\bra{\S}{1...N}{}\sum_{i,i_{1},...,i_{M}=1}^{N}(
\int_{\O_{i}\setminus D^{(R)}}\hat{\F}_{\m}^{(i)}
\int_{\O_{i_{1}}\setminus D^{(R)}\times \cdots\O_{i_{M}}\setminus
D^{(R)}}{\cal R}(\prod_{k=1}^{M}\hat{\F}_{\m_{k}}^{(i_{k})})+\lb
&+&
\int_{{\bf C}\times\O_{i_{1}}\setminus D^{(R)}\times \cdots\O_{i_{M}}\setminus
D^{(R)}}{\cal R}(\hat{\F}_{\m}\prod_{k=1}^{N}\hat{\F}_{\m_{k}}^{(i_{k})})+
[\int_{D^{(R)}}\hat{\F}_{\m}^{(i)},
\int_{\O_{i_{1}}\setminus D^{(R)}\times \cdots\O_{i_{M}}\setminus
D^{(R)}}{\cal R}(\prod_{k=1}^{N}\hat{\F}_{\m_{k}}^{(i_{k})})])=\lb
&=&\bra{\S}{1...N}{}\sum_{i,i_{1},...,i_{M}=1}^{N}(
\int_{\O_{i}\setminus D^{(R)}\times\O_{i_{1}}\setminus D^{(R)}\times 
\cdots\O_{i_{M}}\setminus D^{(R)}}
{\cal R}(\hat{\F}_{\m}\prod_{k=1}^{M}\hat{\F}_{\m_{k}}^{(i_{k})})+\lb
&+&
(\int_{\O_{i}\setminus D^{(R)}}\hat{\F}_{\m}^{(i)}+
\int_{\tilde{\O}_{i}}\hat{\F}_{\m}^{(i)}+
\int_{D^{(R)}}\hat{\F}_{\m}^{(i)})
\int_{\O_{i_{1}}\setminus D^{(R)}\times \cdots\O_{i_{M}}\setminus
D^{(R)}}{\cal R}(\prod_{k=1}^{M}\hat{\F}_{\m_{k}}^{(i_{k})}))=\lb
&=&\bra{\S}{1...N}{}\sum_{i,i_{1},...,i_{M}=1}^{n}(
\int_{\O_{i}\setminus D^{(R)}\times\O_{i_{1}}\setminus D^{(R)}\times 
\cdots\O_{i_{M}}\setminus D^{(R)}}
{\cal R}(\hat{\F}_{\m}\prod_{k=1}^{M}\hat{\F}_{\m_{k}}^{(i_{k})})+\lb
&+&
\int_{{\bf C}}\hat{\F}_{\m}^{(i)}
\int_{\O_{i_{1}}\setminus D^{(R)}\times \cdots\O_{i_{M}}\setminus
D^{(R)}}{\cal R}(\prod_{k=1}^{M}\hat{\F}_{\m_{k}}^{(i_{k})}))=\lb
&=&\bra{\S}{1...N}{}\sum_{i,i_{1},...,i_{M}=1}^{N}
\int_{\O_{i}\setminus D^{(R)}\times\O_{i_{1}}\setminus D^{(R)}\times 
\cdots\O_{i_{M}}\setminus D^{(R)}}
{\cal R}(\hat{\F}_{\m}\prod_{k=1}^{M}\hat{\F}_{\m_{k}}^{(i_{k})})}

Having proven \refbr{nablan} and \refbr{nablansurf} we can move on to determine both 
the curvature torsion and finite parallel transform. 

\subsection{Calculation of the Curvature, Torsion and Finite Deformations}

The curvature tensor ${ R}^{(R)}$ of $\nabla^{(R)}$ is defined by ${ R}^{(R)}(X,Y)
\equiv [\nabla^{(R)}_{X},\nabla^{(R)}_{Y}]-
\nabla^{(R)}_{[X,Y]}$.
When we use the vector fields $X=\*_{\m}$ and $Y=\*_{\n}$ the last
term drops out due to \refbr{derivativescommute}. 
Equation \refbr{nablasquared} therefore implies that 
\eq{{ R}^{(R)}_{\m\n}\ket{\F_{i}}{}{}=
\int_{D^{(R)}\times D^{(R)}}{\cal R}(
\hat{\F}_{\m}\hat{\F}_{\n}-(\m\leftrightarrow \n))
\ket{\F_{i}}{}{}=0\eqn{vanishingcurvature}}
Hence ${ R}^{(R)}=0$.

A connection $\tilde{\nabla}^{(R)}$
on the tangent space is gotten by multiplying  $\nabla^{(R)}$ 
with the projector $\hat{\Pi}$ on the physical subspace,
$\tilde{\nabla}^{(R)}\equiv \hat{\Pi}\na^{(R)}$.
The projector isn't annihilated by $\nabla^{(R)}$ for $R>0$ and hence
$\tilde{\nabla}^{(R)}$ acquires an induced curvature:
\eqs{\tilde{{ R}}_{\m\n}^{(R)}\ket{\F_{\r}}{}{}&=&
\hat{\Pi}\int_{D^{(R)}}\hat{\F}_{\m}\hat{\Pi}\int_{D^{(R)}}\hat{\F}_{\n}
\ket{\F_{\r}}{}{}-
\hat{\Pi}\int_{D^{(R)}}\hat{\F}_{\n}\hat{\Pi}\int_{D^{(R)}}\hat{\F}_{\m}
\ket{\F_{\r}}{}{}
+\lb
&+&\hat{\Pi}\int_{{\bf{C}}\setminus{D^{R}}}\hat{\F}_{\m}\int_{D^{(R)}}\hat{\F}_{\n}
\ket{\F_{\r}}{}{}-
\hat{\Pi}\int_{{\bf{C}}\setminus{D^{R}}}\hat{\F}_{\n}\int_{D^{(R)}}\hat{\F}_{\m}
\ket{\F_{\r}}{}{}}
We see that $\tilde{{ R}}_{\m\n}^{(R)}$ is non-zero for all $R$. However in 
the case of $\d$ we see that both ${R}_{\m\n}(\d)$ and 
$\tilde{{ R}}_{\m\n}(\d)$ are trivially equal to zero.

The torsion $\tilde{{ T}}^{(R)}$ 
of $\tilde{\nabla}^{(R)}$ is defined by
$\tilde{{ T}}^{(R)}(X,Y)\equiv \tilde{\nabla}^{(R)}_{X}Y-
\tilde{\nabla}^{(R)}_{Y}X-[X,Y]$
The torsion vanishes for all $R$: 
\eqs{&{}&\tilde{\nabla}^{(R)}_{\m}\ket{\F_{\n}}{}{}-
\tilde{\nabla}^{(R)}_{\n}\ket{\F_{\m}}{}{}=
\hat{\Pi}\int_{D^{(R)}}(\hat{\F}_{\m}\ket{\F_{\n}}{}{}-
\hat{\F}_{\n}\ket{\F_{\m}}{}{})=0}
where in the last step we have used the
symmetry of on-shell amplitudes:
\eq{\int_{D^{(R)}}d^{2}z\bra{\F_{\r}}{}{}\hat{\F}_{\m}(z,\bar{z})
\ket{\F_{\n}}{}{}=
\int_{D^{(R)}}d^{2}z\bra{\F_{\r}}{}{}\hat{\F}_{\n}(z,\bar{z})
\ket{\F_{\m}}{}{}}

The flatness of $\nabla^{(R)}$ implies that the 
the parallel transport \refbr{finitetrans} $\s_{t_{0}\la t}^{*(R)}:{\cal H}_{t}
\rightarrow{\cal H}_{t_{0}}$
using $\nabla^{(R)}$ is path independent, and we find using \refbr{nablan} and 
\refbr{nablansurf}
\eqs{\s_{t_{0}\la t}^{*(R)}(\ket{\F_{i}}{}{t})&=&
{\cal P}(\exp(\int_{t_{0}}^{t}dt^{\m}\nabla^{t_{0}(R)}_{\m}))
\ket{\F_{i}}{}{t_{0}}=\lb
&=&{\cal R}\exp(\D t^{\m}\int_{D^{(R)}}\hat{\F}_{\m}^{t_{0}})
\ket{\F_{i}}{}{t_{0}}\eqn{deformedextstate}\\
\s_{t_{0}\la t}^{*(R)}\;(\bra{\S}{1...N}{\quad t})&=&\bra{\S}{1...N}{\quad t_{0}}\prod_{i=1}^{N}
{\cal R}\exp(\D t^{\m}\int_{\O_{i}\setminus D^{(R)}}\hat{\F}_{\m}^{(i)t_{0}})
\eqn{deformedsurfacestate}}
It now follows that the parallel transported correlation functions indeed satisfy
\refbr{deformedcorrelator} {\it i.e.}
\eqs{&&\s_{t_{0}\la t}^{*(R)}(\bra{\S}{1...N}{\quad t})\s_{t_{0}\la t}^{*(R)}
(\ket{\F_{i_{1}}}{1}{t})\cdots \s_{t_{0}\la t}^{*(R)}(\ket{\F_{i_{N}}}{N}{t})=\lb
&=&\bra{\S}{1...N}{\quad t_{0}}\prod_{k=1}^{N}{\cal R}\exp(\D t^{\m}
\int_{\O_{k}\setminus D^{(R)}}\hat{\F}_{\m}^{(k)t_{0}}){\cal R}\exp(\D t^{\m}
\int_{D^{(R)}}\hat{\F}_{\m}^{(k)t_{0}})\ket{\F_{i_{k}}}{k}{t_{0}}=\lb
&=&\bra{\S}{1...N}{\quad t_{0}}\prod_{k=1}^{n}{\cal R}\exp(\sum_{\m}t^{\m}
\int_{\O_{k}}\hat{\F}_{\m}^{(k)t_{0}})\ket{\F_{i_{k}}}{k}{t_{0}}=\lb
&=&\vev{{\cal R}(\exp(\D t^{\m}\int_{\S}\hat{\F}_{\m}^{t_{0}})
\hat{\F}_{i_{1}}^{t_{0}}\cdots\hat{\F}_{i_{N}}^{t_{0}})}{\S}^{t_{0}}
\eqn{derivedeformedcorrelator}}
With equations \refbr{deformedextstate}, \refbr{deformedsurfacestate} and 
\refbr{derivedeformedcorrelator} we can in principal, after having shown the sewing 
condition, compare different CFT.
As mentioned in section \ref{marginaldeformation} we can of course express 
the connection $\na_{\m}$ using an ordinary covariant derivative ${\cal D}_{\m}$, which 
for $\na_{\m}^{(R)}$ takes the form 
\eq{{\cal D}_{\m}^{(R)}\vev{{\cal R}(\hat{\F}_{i_{1}}\cdots \hat{\F}_{i_{N}})}{\S}=
\int_{\S\setminus\cup_{i=1}^{N} D_{i}^{(R)}}\vev{{\cal R}(\hat{\F}_{\m}
\hat{\F}_{i_{1}}\cdots \hat{\F}_{i_{N}})}{\S}\eqn{covariantderivative}}
Its clear from the flatness of the connection $\na^{(R)}$ that 
$[{\cal D}_{\m},{\cal D}_{\n}]\vev{{\cal R}(\hat{\F}_{i_{1}}
\cdots \hat{\F}_{i_{N}})}{\S}$ has to vanish. This requires the identity 
\refbr{vanishingcurvature}.

In the next section we will show that the parallel transport obeys the sewing
condition and study the deformation of the conformal structure.

\section{The Sewing Condition and Conformal Invariance}\label{sewingandconformal}

We are now in a position to determine if the sewing condition \refbr{parallelsewing}
is satisfied by the connections defined in section \ref{definitionconnection}.
If we make use of the results in section
\ref{curvature} and use the parallel transform of the reflector
\eq{\s_{t_{0}\la t}^{*(R)}(\ket{R}{ab}{t})={\cal R}exp(\D t^{\m}
\int_{D^{(R)}}(\F_{\m}^{t_{0}(a)}+\F_{\m}^{t_{0}(b)}))\ket{R}{ab}{t_{0}}\eqn{parallelreflector}}
the left hand side of \refbr{parallelsewing} become
\eqs{\s_{t_{0}\la t}^{*}(\bra{\S\infty\S'}{1...N,N+1...N+M}{\quad\quad\quad\quad
\quad\;\; t})=\bra{\S\infty\S'}{1...N,N+1...N+M}{\quad\quad\quad\quad
\quad\;\; t_{0}}\prod_{i=1}^{N+M}
{\cal R}\exp(\D t^{\m}\int_{\O'_{i}\setminus D^{(R)}}\hat{\F}_{\m}^{t_{0}(i)})
\eqn{LHSparallelsewing}}
and the right hand side
\eqs{&&\s_{t_{0}\la t}^{*}(\bra{\S\cup(P_{i},z^{(i)})}{1...N,a}{\quad\quad t})
\s_{t_{0}\la t}^{*}(\bra{\S'\cup(P_{j},z^{(j)})} {N+1...N+M,b}{\quad\quad\quad
\quad\; t})\s_{t_{0}\la t}^{*}(\ket{R}{ab}{t})=\lb
&=&\bra{\S\cup(P_{i},z^{(i)})}{1...N,i}{\quad\quad t_{0}}\bra{\S'\cup(P_{j},z^{(j)})}
{N+1...N+M,j}{\quad\quad\quad\quad\; t_{0}}{\cal R}
\exp(\sum_{i=1}^{N+M}\D t^{\m}(\int_{\O_{i}\setminus D^{(R)}}\hat{\F}_{\m}^{t_{0}(i)}+\lb
&+&\int_{\O_{a}\setminus D^{(R)}}\hat{\F}_{\m}^{t_{0}(a)}+
\int_{\O_{b}\setminus D^{(R)}}\hat{\F}_{\m}^{t_{0}(b)}+
\int_{D^{(R)}}\hat{\F}_{\m}^{t_{0}(b)}+
\int_{D^{(R)}}\hat{\F}_{\m}^{t_{0}(b)}))\ket{R}{ab}{t_{0}}=\lb
&=&\bra{\S\cup(P_{i},z^{(i)})}{1...N,i}{\quad\quad t_{0}}\bra{\S'\cup(P_{j},z^{(j)})}
{N+1...N+M,j}{\quad\quad\quad\quad\; t_{0}}{\cal R}\exp(\sum_{i}^{N+M}
\D t^{\m}\int_{\O_{i}\setminus D^{(R)}}\hat{\F}_{\m}^{t_{0}(i)}+\lb
&+&\int_{\O_{a}\setminus D^{(1)}}\hat{\F}_{\m}^{t_{0}(a)}+
\int_{\O_{i}\setminus D^{(1)}}\hat{\F}_{\m}^{t_{0}(b)})
{\cal R}\exp(\D t^{\m}\int_{D^{(1)}}(\hat{\F}_{\m}^{t_{0}(a)}+
\hat{\F}_{\m}^{t_{0}(b)}))\ket{R}{ab}{t_{0}}=\lb
&=&\bra{\S\cup(P_{i},z^{(i)})}{1...N,i}{\quad\quad t_{0}}\bra{\S'\cup(P_{j},z^{(j)})}
{N+1...N+M,j}{\quad\quad\quad\quad\; t_{0}}{\cal R}\exp(\sum_{i}^{N+M}
\D t^{\m}\int_{\O'_{i}\setminus D^{(R)}}\hat{\F}_{\m}^{t_{0}(i)})\ket{R}{ab}{t_{0}}=\lb
&=&\s_{t_{0}\la t}^{*}(\bra{\S\infty\S'}{1...N,N+1...N+M}{\quad\quad\quad\quad
\quad\;\; t})\eqn{proofsewing}}
The crucial step in \refbr{proofsewing} that remains to be shown is that
\eq{{\cal R}\exp(\D t^{\m}\int_{D^{(1)}}(\hat{\F}_{\m}^{t_{0}(a)}+
\hat{\F}_{\m}^{t_{0}(b)}))\ket{R}{ab}{t_{0}}={\cal R}\exp(\D t^{\m}
\int_{{\bf C}}\hat{\F}_{\m}^{t_{0}(a)})\ket{R}{ab}{t_{0}}=\ket{R}{ab}{t_{0}}\eqn{crucial}}
Actually all we have to do is to saturate the transformed reflector with two arbitrary
external states and use the reflecting property \refbr{reflection} of the reflector and
the regularization \refbr{regularization} or rather its extension 
\refbr{generalregularization}
\eqs{\bra{\F_{i}}{a}{t_{0}}\bra{\F_{i}}{b}{t_{0}}{\cal R}\exp(\D t^{\m}
\int_{{\bf C}}\hat{\F}_{\m}^{t_{0}(a)})\ket{R}{ab}{t_{0}}&=&\bra{\F_{i}}{a}{t_{0}}{\cal R}
\exp(\D t^{\m}\int_{{\bf C}}\hat{\F}_{\m}^{t_{0}(a)})\ket{\F_{i}}{b}{t_{0}}=\lb
&=&\bra{\F_{i}}{a}{t_{0}}\ket{\F_{i}}{b}{t_{0}}=\lb
&=&\bra{\F_{i}}{a}{t_{0}}\bra{\F_{i}}{b}{t_{0}}
\ket{R}{ab}{t_{0}}\eqn{proofcrucial}}

Let's complete the proof of conformal invariance.
Proceeding in the same way as in section \ref{curvature} the finite extension of 
equation \refbr{nablaRVir} is easily shown to be 
\eq{\s_{t_{0}\la t}^{*(R)}\hat{L}_{n}^{t}={\cal R}\exp(\D t^{\m}\int_{D^{(R)}}
\hat{\F}_{\m}^{t_{0}})\hat{L}_{n}^{t_{0}}{\cal R}\exp(
\D t^{\m}\int_{{\bf C}\setminus D^{(R)}}\hat{\F}_{\m}^{t_{0}})\eqn{finiteVirasoro}}
The Virasoro algebra is thus preserved
\eqs{\s_{t_{0}\la t}^{*(R)}([\hat{L}_{n}^{t},\hat{L}_{m}^{t}])=
[\s_{t_{0}\la t}^{*(R)}\hat{L}_{n}^{t},\s_{t_{0}\la t}^{*(R)}\hat{L}_{m}^{t}]=\lb
{\cal R}\exp(\D t^{\m}\int_{D^{(R)}}\hat{\F}_{\m}^{t_{0}})
[\hat{L}_{n}^{t_{0}},\hat{L}_{m}^{t_{0}}]{\cal R}\exp(\D t^{\m}
\int_{{\bf C}\setminus D^{(R)}}\hat{\F}_{\m}^{t_{0}})\eqn{invarianceVirasoroalg}}
In equation \refbr{invarianceVirasoroalg} we have used the identity
${\cal R}\exp(\D t^{\m}\int_{{\bf C}\setminus D^{(R)}}\hat{\F}_{\m}^{t_{0}})
{\cal R}\exp(\D t^{\m}\int_{D^{(R)}}\hat{\F}_{\m}^{t_{0}})=1$.
Even though the conformal weights and the Virasoro algebra 
are preserved, the theory is still deformed since the operator
product coefficients $C_{ijk}^{t}\equiv\bra{\F_{i}}{}{t}\hat{\F}_{j}^{t}(1,1)
\ket{\F_{k}}{}{t}$ are not constant, {\it i.e.}
\eq{\*_{\m}C_{ijk}^{t}=\int_{{\bf C}}\bra{\F_{i}}{}{t}{\cal R}(\hat{\F}_{\m}^{t}
\hat{\F}_{j}^{t}(1,1))\ket{\F_{k}}{}{t}}
does not in general vanish.

\bigskip{\Large\bf Acknowledgment}\medskip

We have benefited from discussions with Martin Cederwall.

\end{document}